# Distributed Sensing, Computing, Communication, and Control Fabric: A Unified Service-Level Architecture for 6G

Dejan Vukobratović[1], Nikolaos Bartzoudis[2], Mona Ghassemian[3], Firooz Saghezchi[4], Peizheng Li[5], Adnan Aijaz[5], Ricardo Martinez[2], Xueli An[3], Ranga Rao Venkatesha Prasad[7], Helge Lüders[6], Shahid Mumtaz[4]

[1]University of Novi Sad, Faculty of Technical Sciences, Novi Sad, Serbia, [2]Centre Tecnològic de Telecomunicacions de Catalunya (CTTC/CERCA) Castelldefels, Spain, [3]Huawei Technologies Duesseldorf GmbH, Munich, Germany*, [4]Instituto de Telecomunicações, University of Aveiro Aveiro, Portugal, [5]Bristol Research & Innovation Laboratory, Toshiba Europe Ltd., UK, [6]Telefonica, Germany, [7]TUD, Netherlands

*Abstract*—With the advent of the multimodal immersive communication system, people can interact with each other using multiple devices for sensing, communication and/or control either onsite or remotely. As a breakthrough concept, a distributed sensing, computing, communications, and control (DS3C) fabric is introduced in this paper for provisioning 6G services in multi-tenant environments in a unified manner. The DS3C fabric can be further enhanced by natively incorporating intelligent algorithms for network automation and managing networking, computing, and sensing resources efficiently to serve vertical use cases with extreme and/or conflicting requirements. As such, the paper proposes a novel end-to-end 6G system architecture with enhanced intelligence spanning across different network, computing, and business domains, identifies vertical use cases and presents an overview of the relevant standardization and pre-standardization landscape.

## I. Introduction

In the 5G era, the mobile communication system performance has improved to another order of magnitude compared to the previous generations, however, the methods to deploy, manage and operate the network have not been improved significantly so far. The management of data and services across different technical and business domains is still very cumbersome, which raises issues like usability, data ownership, security, etc. 5G is designed with strong focus on public network service and utilizes customization as the key to resolve vertical needs. Network slicing is one of the network architecture features introduced in 5G to provide such customization, which is about having virtual networks on top of a public network to serve the industry needs. Network slicing is mainly a Core Network (CN) concept in the 3rd Generation Partnership Project (3GPP) that intends to provide isolated network functions and resources to a large number of subscribers over a shared public network.

The business model of 5G promoted the value of customized Service Level Agreement (SLA) for industry customers. Mobile operators have good experience in providing SLA for public services. However, SLA and corresponding management schemes for vertical sectors are new topics in the telecommunications domain. The 5G Alliance for Connected Industries and Automation (5G-ACIA) have shared their view on guaranteed Service Level Specification (SLS) i.e., the technical part of an SLA [1], which specifies the vertical needs on the SLS and the gaps that should be addressed by future mobile communication systems. Within the industrial domain, vertical customers often prefer data to remain within their premises at a local (edge) cloud for reasons beyond performance requirements such as data privacy and security. It is foreseen that edge computing will be a major trend for the next decade. To this end, a challenge would be to interconnect all distributed resources at the edge at large scale instead of having many isolated data islands.

For 5G and 5G-Advanced, it is envisioned that it will span across multiple releases, including Release 18, Release 19, and perhaps Release 20 and beyond and still has room to improve to fully embrace the requirements of different vertical industries. Even though gradual improvements such as the network slicing in Radio Access Network (RAN) advancements are promised by 5G standardization, radical system architecture changes are expected to better respond to capture vertical needs to meet a true end-to-end (e2e) vision with guaranteed performance for different services, ranging from edge computing to Industrial Internet of Things (IIoT) and integrated sensing and control. At this point in time, with the current picture of 5G research and standardization progress and the feedback from commercial adaptation, especially from vertical industry sectors, it is clear that there is a need to move towards a new system architecture, which could address the discussed gaps and challenges through an innovative design. This requires a holistic research approach towards the needed technology, with a value chain perspective.

From a comprehensive system perspective, the target is to develop a unifying network architecture featuring orchestration, and Artificial Intelligence/Machine Learning (AI/ML) models and frameworks to offer distributed and reliable sensing, communications, computing and control services over 6G. In this paper, we introduce a new innovating 6G architecture built based on a proposed Distributed Sensing, Computing, Communication, and Control (DS3C) fabric, which will serve as a distributed intelligent network that will craft the real-time interactions between human, physical, and digital worlds, helping to set up a new era in which everything could be sensed, connected, and controlled.

## II. The DS3C Motivation and Concept

Considering the huge business opportunities in industrial automation, healthcare and consumer market use cases (UC), a unified communication service to deliver data and support control and actuation is required. Therefore, it is essential to design a 6G system architecture that could tackle such requirements from a high-level perspective, referred to as intents, and transform them into low-level needs, i.e., resources to be allocated leading to meeting the service KPIs/SLAs. For example, a sub-millisecond latency is demanded for services based on human haptic perception in collaborative robots (cobots), as well as long range remote operation which are currently not supported by 5G systems. Moreover, the heterogeneity of services imposes a big challenge to handle with a unified service provisioning system. UCs with extreme requirements can be categorized based on their interaction types between human, digital, and physical worlds, and the service area: local interactions e.g., between mobile cobots and operators on a factory floor;

remote operation such as tele-surgery; and meta-interaction such as collaborative design between engineers in different locations over metaverse.

The role of the DS3C fabric is to coordinate distributed resources in an intelligent manner, to sense and collect physical world information via a communication, sensing, and computing platform and utilize them especially for control domain as shown in Fig. 1. In this way, the physical world information could be managed in the virtual world (e.g., Digital Twin) and the knowledge that is extracted from such information could be used by the physical world. In DS3C, a control loop is closed via the common communication network which multiplexes digital data from the RF or non-RF sensors to the controller and from the controller to the actuator along with the multimodal traffic from other control loops and management functions. Distributed computing and AI/ML will enable accurate sensing and modelling over unique user interfaces essential for the computing system to communicate with the controller and with interfaces for executing the actuation commands.

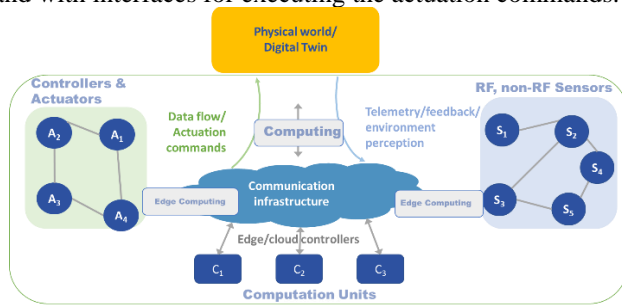

Fig. 1: The main DS3C fabric components

The DS3C focuses on provisioning key features, which will enable enhanced intelligent management of mobile communication, computation, and control and automation resources, natively incorporating intelligent algorithms across the 6G network as highlighted in the following:

- Prediction and monitoring of service and network parameters, and intelligent resource management across network domains in a holistic e2e manner.
- Integration of sensing, ultra-reliable low latency communications (URLLC), and edge computing to support vertical industries with extreme performance requirements (e.g., cobots and autonomous vehicles).
- Automatic discovery and abstraction of the underlying DS3C resources, including communications, computation, sensing and actuation.
- Integration with AI-empowered framework to assist the network orchestration and resource-related decisions, e.g., exploiting the trained models for resource provisioning and optimization, QoS prediction, anomaly detection, and energy efficiency.
- Mapping the extreme performance requirements of the verticals SLAs into low-level needs at both service KPIs (e.g., service deployment time, maximum permitted e2e latency) and resources needs (e.g., guaranteed bandwidth and average utilization of computing resources).
- Selecting and using the tools/mechanisms/solutions to retrieve monitoring, telemetry, and sensing information.
- An e2e service provisioning and reconfiguration using a holistic cross-domain AI-based resource manager that reacts quickly to dynamic service requests.
- Fine tuning of fast resource management decisions considering domain-specific resource management modules that will iteratively optimize decisions until consensus is reached across all participating domains.

### III. CHALLENGES AND OPPORTUNITIES

The proposed DS3C fabric contributes addressing current challenges for an e2e service delivery as detailed in the following.

#### A. Unified Network Orchestration

To reduce or ideally eliminate the manual intervention, improving autonomous management capabilities and functionalities is needed. This can be done based on closed-loop automation and AI/ML algorithms for predictions and decision making, particularly for provisioning, monitoring, self-organization/optimization, self-healing, and performance assurance. These constitute the pillars to attain effective e2e service lifecycle management over cross-domain and multi-technology infrastructures as defined by ETSI Zero-Touch Service Management (ZSM) [2].

*DS3C contribution:* The proposed orchestration platform relies on a hierarchical approach: i) the higher-layer orchestrator interacts with the verticals (mapping intents/high-level requirements to low-level requirements/KPIs) and handles the service lifecycle (e.g., planning, optimization, and SLA assurance) from an e2e perspective. This higher-layer element coordinates multiple domains/technologies (i.e., sensing, computing, networking, and control) to configure the underlying resources. ii) the lower-layer entity is formed by a pool of dedicated domain controllers bound to a specific technology/segment (e.g., computing, sensing, RAN, transport). The DS3C fabric aims to provide APIs between the building elements to achieve the autonomous functions, and also defines a data modelling which abstracts the heterogeneous attributes and capabilities of every domain.

#### B. Native Support for AI/ML in 6G

AI/ML models can be used to solve complex or currently intractable problems for handling network management tasks (e.g., traffic forecasting, anomaly detection, and resource allocation), paving the way towards zero-touch network and service management systems [2]. AI/ML could be used in different layers to adaptively optimize and enhance resource utilization to further push the performance boundaries in cellular systems. For instance, at radio link layer, resource scheduling or link adaptation can be better implemented through radio channel or user equipment location prediction. At network layer, AI/ML can be used for traffic prediction and anomaly detection. 5G has begun to integrate AI/ML in its architecture design, e.g., by introducing the Network Data Analytics Function (NWDAF) [3] to 5G Core for the automation and optimization of the related network functions (e.g., AI/ML-based mobility management). The role of NWDAF function is to collect and analyze data from other 5G network elements to train AI/ML models that can be used by network services. Meanwhile, a similar mechanism like collecting and analyzing data based on the existing Self-Organizing Networks and Minimization of Drive Tests (SON/MDT) is adopted for 5G radio access networks (RAN).

*DS3C contribution:* The emergence of NWDAF is only the starting point to support AI/ML. It is an add-on approach to NWDAF that could be placeholder for AI/ML related purposes, as well as the interfaces among the relevant network entities that may interact with NWDAF. In DS3C, a clean slate approach could be adopted to define a unified system architecture, by providing native support for AI/ML inclusion to facilitate data governance and system orchestration for a better management of converged infrastructure resources with a focus more on architectural aspects rather than functional aspects.

### C. AI/ML Models for Resource Management

Current research on AI-based RAN/CN architecture is decoupled into two main streams: 1) definition of architectural frameworks covering aspects such as definition of modules and interfaces, and problems of data collection, model training, deployment and life cycle management, and 2) AI/ML architectures suitable for different, usually intra-domain resource management tasks. Examples open-source communities are ETSI OSM and ETSI ONAP that have delivered initial open-source management and orchestration implementations in 5G.

***DS3C contribution:*** Establishing a fast and hyper-reactive e2e service across multiple domains requires fast and accurate decisions to be available at the highest level of network management architecture almost instantaneously, which is feasible using suitable AI/ML models combined with flexible and programmable infrastructure. Intelligent resource management will require novel data sources to be included in resource management decisions. Sensing capabilities of devices and infrastructure, channel state information including the existence of reconfigurable intelligent surfaces (RIS), beam configurations and management, and precise localization information will provide informative inputs to future resource management algorithms. Besides data sources, the question of energy efficient AI/ML model training, robust and dependable AI model inference, and economic life-cycle management of AI models are key aspects of the DS3C intelligent fabric design. The goal is to minimize the required data consumption while selecting models with minimal parameter and training complexity, and helping the selection of robust explainable AI models.

### D. Integrated Sensing and Communications (ISAC)

It is widely believed that 6G will be designed for simultaneous communication and sensing, e.g., by exploiting the imaging capability of radio frequency (RF) signals in mmWave/THz bands and the abundance of radio spectrum in these bands to transmit high resolution Radar and Ultra-Wide Band (UWB) wireless localization signals using Frequency Modulated Continuous Wave (FMCW) signals and/or extremely short pulses. Among others, this functionality can enable an intense use of virtual or augmented reality applications, e.g., in smart cities/factories, using contextual information (e.g., user positioning and/or object detection) provided by the sensing fabric.

***DS3C contribution:*** The DS3C fabric advocates the integration of sensing functionality into the mobile communications and computing (edge/cloud) infrastructure to provide not only real-time telemetry for monitoring, predicting, and resource management in 6G (e.g., channel estimation and beam management), but also sensing services (e.g., wireless localization) to the vertical industries (e.g., to accurately position human, objects, and robots in smart factories and monitor human activities for human safety or healthcare applications). Sensing can be integrated in the DS3C fabric at different levels. At high level, the sensing and communication systems can be separated, and the continuous information exchanged between them can ensure an improved spectrum sharing. On the other hand, at physical or medium access control layer, sensing and communication signals can be multiplexed in time, frequency, and space, enabling the two systems to share not only the spectrum but also the hardware.

Apart from the regular sensing nodes and IoT sensors, the DS3C fabric aims at leveraging a crowdsourcing solution to deliver an ISAC framework at the network edge, which will be based on active sensing (e.g., beam scanning), passive sensing (e.g., channel sounding) and interactive sensing (e.g., via data transmission). Such paradigm could be used in multiple UCs such as area imaging, activity recognition, spatial-aware computing, localization and mapping, and RF maps e.g., for the metaverse.

### E. Converged Communications and Computing Networks

The edge computing concept has emerged in parallel with the development of mobile communication systems. It could be considered as a distinguished form of cloud computing that moves part of the processing and data storage resources away from the central cloud to the edge cloud located physically and logically close to the data providers and data consumers [5]. ETSI has been investigating this topic under its Multi-access Edge Computing (MEC) Industry Specification Group (ISG) for mobile communication systems [6]. Moreover, MEC features the Radio Network Information Service (RNIS) interface, which exposes various RAN information to the Edge platform though a dedicated API which is the focus activity in 3GPP SA6 group.

***DS3C contribution:*** The DS3C fabric aims to bring the edge computing concept to a new level, enabling the so called "edge-cloud continuum", where edge and cloud are increasingly undistinguishable (i.e., used homogeneously by the service provision layer). 6G will employ this edge-cloud computing fabric to enable distributed and serverless computing for AI/ML applications exploiting, e.g., federated learning. This will unify distributed communication and computing resources from the access to transport and core, which will break the conventional physical boundaries of technology domains. Moreover, novel switching, resource run-time scheduling and management mechanisms that take all dimensions as well as constraints into account, will be inherent building blocks of the DS3C fabric. The converged DS3C fabric will address edge-specific requirements originating especially from vertical UCs, e.g., connecting resource constrained IIoT nodes with extreme and conflicting performance requirements.

### F. Communication and control co-design

The existing work on communications and control co-design can be divided into two main categories: (1) "control of networks" (2) "control over networks" [7]. While the first category focuses on satisfying the QoS requirements of Networked Control Systems (NCSs) from the communication network's perspective, the second aims at mitigating the performance loss due to the network's limitations within the application layer. In essence, the key idea of communication and control co-design is to consider the dynamics of the control system as well as the network's characteristics jointly to improve the overall performance of the system. This is different from the conventional methods and trends in networking research, as the primary goal is not to maximize throughput or minimize latency but rather to guarantee stability or maximize control performance.

*DS3C contribution:* The DS3C will embrace recent trends on communications and control co-design which proposes KPIs beyond the conventional ones used in today's communication networks such as age of information (AoI), considers measuring information freshness in remote monitoring and control scenarios [7], age of incorrect information (AoII), urgency of information (UoI) as well as Value of Information (VoI) to quantify the application layer performance, as an alternative to latency, throughput or jitter. Each of these metrics captures a different aspect of control performance and offer great potential for future industrial networks. Furthermore, for UCs with actuation and control requirements, communication–control co-design methods require strong correlations between control and communication systems, i.e., control optimization problem with communication constraints and communication optimization problem with control constraints. To apply the communication–control co-design, the essence of the co-design problem should build an effective co-design model to help with communication and control optimization.

## IV. THE DS3C-BASED 6G ARCHITECTURE

By integrating sensing, computing, communication, and control capabilities, the DS3C fabric will be able to sense and process information in real-time, make intelligent decisions based on that information, and adapt to changing network conditions to optimize performance. Composed of four hierarchical stratums: network, control and actuation, network intelligence, and sensing stratum, the high-level architecture that relates DS3C fabric and strata is presented in Fig 2.

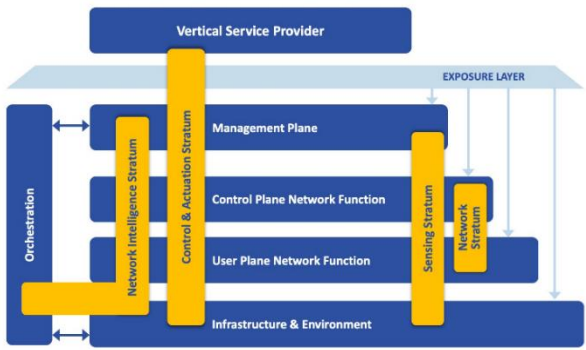

Fig 2: The DS3C strata

In summary, the optimized automation of the DS3C fabric will be based on three technical pillars, which can be briefly summarized as: i) AI-based prediction and monitoring (thereafter AI-PM) trained models integrated in the DS3C fabric at different hierarchical levels to serve resource management at local and global level, ii) AI-driven holistic resource management (thereafter AI-HRM as shown in Figure 5) aiming at the coarse resource allocation vertically, across different domains, and horizontally, across different edge sites, iii) AI-driven domain-specific resource management (thereafter AI-DRM) covering the sensing, communication and compute domains and sharing information and policies with the holistic resource management framework. The key components of the proposed DS3C-based architecture are depicted in Fig. 3 and summarized below.

### A. Edge platform

Adopting an existing, preferably open framework (e.g., ONF's Aether, Intel's Smart Edge Open, OpenNebula) facilitates the configuration of control plane functions and edge node management services (e.g., virtualization infrastructure, container runtime environments). The virtualization capabilities of the Edge platform will be extended to support the partitioning of accelerated functions in different co-processing elements (i.e., GPUs, FPGAs, Superscalar processors). The edge platform will also interact with the data collection and pre-processing framework.

### B. Two-level orchestration layer

Local site orchestration is used to leverage the complete compute, memory, and I/O blueprint of the site to be able to optimally exploit the allocation of resources in the underlying heterogenous hardware substrate. Towards this end, current orchestration solutions will have to be extended in the context of the DS3C fabric, to introduce the notion of specialized micro-orchestrator modules, controlling specific physical or logical computing entities of the local site and able to apply field partitioning of the computational workloads and tasks (e.g., covering both application and network functions). The local site orchestrator will also feature ETSI MEC-compliant interfaces and establish bi-directional connectivity with other control entities at RAN and transport levels. The automation operations will be achieved adopting closed-loop through the AI-DRM pillar. This can make use of (integrated) open-source solutions like the "Eclipse fog05", ETSI TeraFlowSDN, ETSI OSM, that will provision and manage decentralized compute, storage, communication, and control resources across the network infrastructure.

*Cross-site unified orchestration* will provision and deploy the e2e service orchestration covering multiple edge sites and considering the entire edge-to-cloud compute continuum. The closed-loop automation will be achieved through the AI-HRM and by integrating distributed AI solutions and data-intensive workflow task schedulers. Moreover, a custom interface with the e2e slice manager will be built in 6G to enable the cross-domain implementation of joint computing and network resource allocation and policy distribution, satisfying SLAs, KPIs and UC specific requirements.

### C. Data collection and pre-processing

This module leverages different monitored metrics across the DS3C fabric, exposed by the sensing, the computing infrastructure (i.e., the edge platform and the orchestrator), the network and the application domain. This layer will feed the AI-PM module that is described in the following subsections.

### D. AI Prediction and Monitoring (AI-PM) module

AI-PM module will host AI models for AI prediction and monitoring (anomaly detection) and will be flexibly instantiated and deployed across different domains and locations. Both training the AI models and their deployment for inference tasks will be done at AI-PM module based on the data retrieved from the Data collection and Pre-processing module. AI models hosted by the AI-PM module will predict main system parameters: system load, the resource usage and KPIs. These predictions will be forwarded to resource management modules AI-HRM and AI-DRMs for further decision making. Predictions will be made based on incoming data and AI models for monitoring and detecting anomalies in case the system state and KPIs are deviating from the expected behavior. Information on detected anomalies will be sent to the AI-HRM and AI-DRMs for decision making. The AI-PM will host AI models for different prediction and monitoring tasks, each of which may exploit different deep learning (DL) architectures and approaches. The pre-trained deep neural networks will be applied along with the use transfer learning algorithms to fine-tune non-stationary 6G mobile networks, likewise improving learning efficiency and performance. When deployed in scale, AI-PM predictors can use federated learning to improve their performance.

tasks for radio resource allocation, beam management, RIS management and control, resource management and control related to sensing and spectrum sharing. Multi-task learning with DNNs and DRL will be explored to aggregate activities on different resource management tasks, produce more efficient and lower footprint models. Resource management DL approaches will be based on data-based constrained DRL methods and model-based active inference approaches.

*F. Holistic Resource Management (AI-HRM)*

This is a central module that will manage e2e resource allocation in a holistic manner. It will receive inputs from AI-

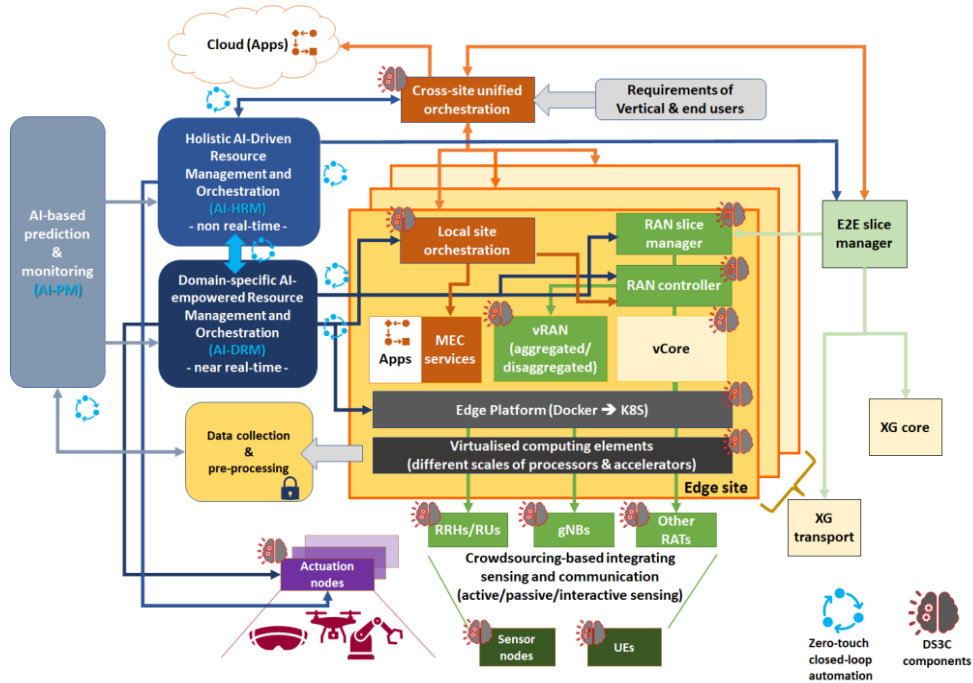

Fig 3. Proposed 6G architecture based on the DS3C fabric

*E. Domain-specific Resource Management (AI-DRM)*

Domain-specific RM modules will be deployed across the central and edge sites to support decision making within different network domains. AI-DRM modules will receive inputs from AI-PM modules in the form of data streams, predictions, and anomaly events. These inputs will feed AI models hosted within the AI-DRM modules. AI models will be trained at the AI-PM module and deployed for inference tasks at the AI-DRM module to avoid significant data flows among many deployed AI-PMs and AI-DRMs. Due to variety of resource management tasks at different domains, careful design of AI models, their interaction and data sharing within each AI-DRM module will be needed and this will constitute one of the main challenges of AI-DRM design. In other words, careful balance between simple AI models targeting specific tasks and more complex multi-task AI models will be explored as AI-DRM architectural choices. The outputs of the AI-DRM module will feed to various network controllers and orchestrators (e.g., RAN controller, edge platform orchestrator) for resource allocation decisions. These decisions will be updated through interaction with AI-HRM to reach a consensus on e2e resource allocations.

AI-DRM will host a rich collection of different AI models for domain-specific and task-specific resource management. Examples include a collection of radio resource management

PM in the form of data streams, predictions and anomaly events and will trigger resource management decisions at the highest hierarchical level. The AI-HRM will respond to dynamically changing vertical service intents and network anomalies and for each such inputs a two-phase procedure will follow. In the first and time-critical phase, AI-HRM will react with resource management decisions that provide best high-level response to the received inputs. In this way, service continuity and provisioning of requested SLAs will be maintained with high resilience. In the second phase, information exchange through distributed message-passing between AI-HRM and relevant AI-DRMs will follow, where resource allocation optimization will evolve through consensus-based methods.

The AI-HRM module will host AI models for RAN, transport, CN slicing decisions that receive inputs from the AI-PM module in the form of data streams, predictions and anomalies. It will exchange information with the AI-DRM modules during the resource optimization (hardening) phase. Responsive AI models for initial and fast network (re)configuration will use GNN-based models and constrained-based DRL models. Joint system optimization between AI-HRM and selected subset of AI-DRMs will rely on consensus-based DRL models.

## V. STANDARDISATION ROADMAP

5G introduced as a native service-based architecture (SBA) [3GPP23.501] especially for the CN to support three main UCs: enhanced mobile broadband (eMBB), URLLC and massive machine-type communications (mMTC). Benefiting from the cloud and virtualization technologies, the SBA in 5G is capable of dramatically reducing the specification complexity, especially on relevant network functions, interfaces and procedure definitions for newly introduced services. Many new features (e.g., cellular IoT and URLLC) [3GPP23.501] are added on top of the design of network services provisioned for public network services with joint efforts of 3GPP SA1, SA2, SA3, SA4, SA5 and SA6 from Release 15, 16, 17, 18 to Release 19.

Such initiatives could be considered the first standardization attempt to natively support edge computing in 6G. It is expected that DS3C-related standardization activities will take place in parallel, in multiple SDOs, and possibly in multiple stages; initially in the form of consensus on 6G vertical requirements and architectural design, and later through harmonization of computing, communication and control and their interfaces and APIs.

Currently, two 3GPP Working Groups (WGs) are active on these topics. Key vertical UCs (especially for factory and process automation) are studied in 3GPP System Aspects (SA) WG1 (SA1) [3GPP22.104] and SA6, [3GPP TR 23.745], whereas since 2021 the 3GPP also approved to investigate architectural enhancements for Extended Reality and media (XRM) services [3GPP26.928], which are also relevant to the verticals. SA2 takes requirements specified in SA1 such as [3GPP22.847] on tactile and multi-modal communication services for enhancing 3GPP system functionalities. Performance requirements for AI/ML Model Transfer (AMMT) in 5GS is amongst the relevant Rel-18 features that will pave the way for adding support to metaverse services in Rel-19. In this respect, the combination of haptics type of XR media and other non-haptics types XR media are expected to be analyzed, identifying users and digital representations that interact within metaverse services, together with physical and digital information collected and used to enable metaverse services. The feasibility study on localized mobile metaverse services, wherein the potential metaverse engineered UCs and requirements are summarized [3GPP22.856].

Other than 3GPP, the ITU's Focus Group on Machine Learning for Future Networks including 5G [FG-ML5G] has also drafted several technical specifications for ML for future networks, on topics such as interfaces, network architectures, protocols, algorithms, and data formats [4]. Another related SA activity is the Metaverse focus group [MV-FG] that aims to analyze the technical requirements of the metaverse to identify fundamental enabling technologies in areas from multimedia and network optimization to digital currencies, Internet of Things, digital twins, and environmental sustainability.

## VI. SUMMARY AND OUTLOOK

The proposed DS3C fabric is key to the 6G development at providing a unified, distributed, and agile framework to enable data pipelines coming from the entire sensing stratum. The DS3C articulates its operation by applying opportunistic data caching and pre-processing, and host the execution of high volume and highly complex computational workflows across a 6G-enabled and connected infrastructure of heterogenous computational resources that includes distributed endpoints (e.g., embedded devices) with sensing and actuating capabilities, edge platforms (e.g., edge/cloud computing servers, fog nodes, micro-datacenters), and remote cloud datacenters.

To that end, the DS3C will architect a flexible and scalable edge processing design, blending the lifecycle management and closed-loop automation for cloud-native network functions, edge applications, and network services across a multi-level edge-to-cloud continuum. To achieve that, an underlying processor-agnostic framework will fully exploit the capabilities of general-purpose processors, and bit-intensive acceleration processing and communication elements, able to serve challenging processing loads and ultra-low actuation latency to meet the extreme use case KPIs.

ACKNOWLEDGMENT

*DS3C is a concept developed by: Toshiba, Huawei, UNS, CTTC, ITAV, RWTH, TEF, TUD, Martel, Siemens, IRIS as part of SNS Horizon JU programme SNS SCALAR-6G proposal submitted in April 2023.*